\documentclass[
    a4paper,
    12pt,
    amsmath,
    amssymb,
    prc,
    superscriptaddress,
    floatfix,
    longbibliography
]{revtex4-1}
\usepackage[utf8]{inputenc}
\usepackage{graphicx}
\usepackage{hyperref}
\usepackage[english]{babel}

\begin{document}

\title{Is it possible to simulate the cosmological stage of the nucleon mass generation?}
\author{Vladimir I.\ Komarov}
\email{Email: komarov@jinr.ru}
\affiliation{Joint Institute for Nuclear Research, RU-141980 Dubna, Russia}
\date{\today}
\begin{abstract}
The mechanism of elementary particle mass generation is a fundamental issue of physics.
While cardinal progress has recently been achieved in the case of leptons and current quarks through the Brout-Engler-Higgs mechanism, this issue is at the stage of searching for the constituent quarks forming the nucleon mass.
A deep similarity is noted in the paper between the constituent quark states arising in central inelastic $NN$ collisions within a certain energy interval and the QCD matter states at the cosmological stage of nucleon mass generation.
This gives rise to special attention to the investigation of both issues.
\end{abstract}
\maketitle

The mechanism of elementary particle mass generation is a fundamental problem of science.
The discovery of the Brout-Englert-Higgs mass generation mechanism~\cite{Englert1964,Higgs1964} triumphantly confirmed by experiments, turned out to be the largest achievement of particle physics of the last decades.
The Higgs boson mechanism generates masses of leptons and current quarks but is not relevant to the nucleon mass generation (see e.g.~\cite{Roberts2020}).
However, it is this mass that makes up at least 80\% of the Universe mass in the measurable components, leptons and hadrons, and an even greater part on the Earth.

A reliable and generally accepted approach to the description of nucleon mass generation does not currently exist.
It is only accepted that this generation is due to spontaneous violation of the vacuum chirality symmetry, but the type of transition mechanism of ``light'', 5-MeV, current quarks to ``heavy'', 300-MeV constituent quarks forming the nucleons remains unknown.
The following brief scenario of this transition may be assumed in cosmology.
The cooling of matter in the primary quark-gluon phase led to a decrease in the momentum of current quarks down to a value of the order of tens of GeV.
This increased the QCD interaction constant up to the value at which the interaction of the quark with the QCD vacuum generated gluon correlations of about 0.3~fm in size~\cite{Giacomo1992,Kopeliovich2007} around the quark.
This absorbed its kinetic energy with a corresponding increase in the quark mass: a current quark turned into a constituent quark.
The latter stayed to be colored like the current one.
The space became filled with such quarks exchanging gluons and their correlations, Goldstown bosons.

Such a phase of the QCD matter can be called the \textit{constituent-quark phase} (CQP).
An important characteristic of this phase is the formation of nucleons during its existence.
One may assume formation, first, of diquarks in $qq$ collisions with subsequent diquark-quark impacts.
The latter should be accompanied by a contact with gluon field units of a specific kind providing for the $qqq$ binding of the colorless baryon structure.
It resulted ultimately in the production of nucleons carrying the main part of the universe mass.

The need to study and simulate the discussed CQP is obvious.
Unfortunately, all this extremely important stage does not attract attention of modern researchers: both experimental and theoretical efforts are concentrated exclusively on the study of the hadron and quark-gluon phases of the QCD matter.
Huge efforts are made to simulate the quark-gluon phase using heavy-ion collisions at energies of hundreds of GeV and higher.

Recently, attention was attracted to the area earlier bypassed by studies, \textit{nucleon-nucleon inelastic central collisions} (ICC) at intermediate energies of several GeV to tens of GeV~\cite{Komarov2018,Komarov2022}.
These collisions with impact parameters less than about 0.4~fm at c.m.s.\ energies $\sqrt{s} > 5$~GeV result in an overlap of the central quark content of the nucleons.
Such a collision produces an intermediate excited state free of any baryons.
The valence constituent quarks of the initial nucleons not destroyed in the process of mutual braking of nucleons are conserved and interact through the exchange of gluons and their correlations.
The baryon number $B = 2$ is stored in the system through a special baryonic component of the gluonic field~\cite{Garvey2000}.
The system expands and cools down to an average quark momentum $Q$ of the order of $L_\mathrm{QCD} = 0.33$~GeV when hadronization starts and two nucleons are reestablished for the final state.

Destruction of the constituent quarks at the first stage of the collision occurs only at c.m.s.\ energies providing for a quark momentum higher than about $L_\chi$, the momentum of the chirality restoration.
This quantity was estimated to be at the level of 1~GeV already as far back as in the 1980s~\cite{Manohar1984}, but is still not determined in experiments and may be expected to be somewhat higher.
Therefore, the central collisions should be accompanied by the formation of constituent quark systems with quark momenta in the interval
\begin{equation}
    L_\mathrm{QCD} < Q < L_\chi.
    \label{Qlim}
\end{equation}

The QCD state of the constituent quarks surrounded by the gluonic field can be termed the \textit{constituent quark state} (CQS) quite similar to the states in the cosmological constituent quark phase, CQP, considered before.
Therefore, the relevant investigations of CQS in central $NN$ collisions and of CQP at the cosmological stage of the nucleon mass generation have quite general goals.

Realization of CQS could be achieved~\cite{Komarov2018,Komarov2022} as the intermediate states in reactions of the type
\begin{align}
    N + N &\to N(90^\circ) + N(90^\circ) + M, \label{plus90} \\
    N + N &\to N(90^\circ) + N(-90^\circ) + M, \label{minus90}
\end{align}
where $N(90^\circ)$ denotes the nucleon emitted orthogonally to the reaction axis and $N(-90^\circ)$ is the nucleon with a momentum directed opposite to that of $N(90^\circ)$.
$M$ is the system of light mesons, predominantly pions and also pairs of kaons.
Less probable hyperon and nucleon-antinucleon pair production can be ignored in a first approximation.

The simplest for measurements and analysis may be one of the reactions like~\eqref{plus90}:
\begin{equation}
    p + p \to d(90^\circ) + M.
    \label{dM}
\end{equation}
It is important that their expected cross sections are sufficiently large for experiments with modern proton beams.
The differential cross section summed over all channels is evaluated~\cite{Komarov2022} at about 1~$\mu$b/sr and is expected to change incidentally in the range
\begin{equation}
    5~\mathrm{GeV} < \sqrt{s} < 10~\mathrm{GeV}.
    \label{Elim}
\end{equation}
This approximately definite energy range corresponds to conditions~\eqref{Qlim}.

The main reason for the expected small variation of the cross section is the corresponding slow change of the ICC cross section closely related to the geometrical cross section of the nucleon core.
A similar feature of the high energy hadron-hadron collisions with multiple meson production as a sign of centrality was noted a long time ago by R.~Feynman~\cite{Feynman1969}.

Reactions (\ref{plus90}--\ref{dM}) open up wide opportunities for experimental studies of momentum and polarization distributions of the observables and their correlations in different reaction modes.
An important issue, also, is the $\sqrt{s}$ dependence of these distributions.
The data to be obtained should definitely carry information about the properties of the constituent quark states (CQS).
It is completely unsatisfactory that at present there are no any phenomenological models predicting the values of these specific quantities to demonstrate the CQS properties from experiments.
So, a typical task is to prepare search experiments and relevant theoretical models for studying inelastic central $NN$ collisions at energies~\eqref{Elim}.
As I tried to show above, this activity can concern significant problems of the QCD matter as well as simulations quite similar to the widely used simulations of the quark-gluon cosmological stage through heavy-ion high energy collisions.

\begin{acknowledgments}
I am grateful to the participants of scientific seminars in the JINR Bogolyubov Theoretical Physics Laboratory and Dzelepov Laboratory of Nuclear Problems for useful discussions.
\end{acknowledgments}

\bibliography{cosmological}

\end{document}